\documentclass[12pt]{article}%
\usepackage{cite}
\usepackage{fix-cm}
\usepackage{color}
\usepackage[usenames,dvipsnames]{xcolor}
\usepackage{heck}
\usepackage{amsmath}
\usepackage{amsfonts}
\usepackage{amssymb}
\usepackage{bbold}
\usepackage{graphicx}
\usepackage{multicol}
\usepackage{hyperref}
\usepackage{setspace}
\usepackage[all]{xy}
\usepackage{verbatim}
\usepackage{color}
\usepackage{epsfig}
\usepackage{mathtools}
\usepackage[margin=1in]{geometry}

\providecommand{\U}[1]{\protect\rule{.1in}{.1in}}
\def\bZ{\mathbb{Z}}

\newcommand{\cG}{\mathcal{G}}

\newcommand{\cN}{\mathcal{N}}

\def\be{\begin{equation}}
\def\ee{\end{equation}}
\def\bea{\begin{eqnarray}}
\def\eea{\end{eqnarray}}
\def\spc{\hspace{1pt}}

\def\zz{\hat{z}}
\def\ww{\mu}
\newcommand{\smpc}{\hspace{.5pt}}
\def\is{ \! &  \! = \!  & \!   }
\def\ra{\rangle}
\def\la{\langle}
\def\nspc{\hspace{-.5pt}}

\def\half{\frac {\raisebox{-1pt}{\footnotesize 1}} {\raisebox{1pt}{\footnotesize 2} }}

\definecolor{gray}{rgb}{0.5,0.5,0.5}
\def\AB{{\mbox{\nspc\nspc\fontsize{7pt}{.05pt}$A\nspc\smpc B$}}}
\def\AA{{\mbox{\nspc\nspc\fontsize{7pt}{.0pt}$A$}}}
\def\BB{{\mbox{\nspc\nspc\fontsize{7pt}{.05pt}$B$}}}
\def\AAn{{\mbox{\nspc\nspc\fontsize{8pt}{.0pt}$A$}}}
\def\BBn{{\mbox{\nspc\nspc\fontsize{8pt}{.05pt}$B$}}}

\def\li{{\bigl |}}
\def\ra{{\bigr\rangle}}
\def\la{{\bigl\langle}}
\def\ABCDE{{\mbox{\nspc\fontsize{7pt}{.05pt}$ABCDE$}}}
\def\CD{{\mbox{\fontsize{7pt}{.05pt}$CD$}}}
\def\AD{{\mbox{\nspc\fontsize{7pt}{.05pt}$AD$}}}
\def\AC{{\mbox{\nspc\fontsize{7pt}{.05pt}$AC$}}}
\def\BD{{\mbox{\fontsize{7pt}{.05pt}$BD$}}}

\def\BC{{\mbox{\fontsize{7pt}{.05pt}$BC$}}}

\def\CC{{\mbox{\fontsize{7pt}{.05pt}$C$}}}

\def\half{\frac {\raisebox{-1pt}{\footnotesize 1}} {\raisebox{1pt}{\footnotesize 2} }}

\definecolor{gray}{rgb}{0.5,0.5,0.5}
\def\AB{{\mbox{\nspc\nspc\fontsize{7pt}{.05pt}$A\nspc\smpc B$}}}
\def\AA{{\mbox{\nspc\nspc\fontsize{7pt}{.0pt}$A$}}}
\def\BB{{\mbox{\nspc\nspc\fontsize{7pt}{.05pt}$B$}}}
\def\AAn{{\mbox{\nspc\nspc\fontsize{8pt}{.0pt}$A$}}}
\def\BBn{{\mbox{\nspc\nspc\fontsize{8pt}{.05pt}$B$}}}
\begin{document}
\def\iimath{\dot{\imath}}
\addtolength{\abovedisplayskip}{.65mm}
\addtolength{\belowdisplayskip}{.65mm}
\addtolength{\baselineskip}{.5mm}
\addtolength{\parskip}{.5mm}
\renewcommand{\footnotesize}{\small}

\date{January 2014}

\title{\LARGE   
Covariant Non-Commutative Space-Time}

\institution{HarvardU}{\centerline{${}^{1}$Jefferson Physical Laboratory, Harvard University, Cambridge, MA 02138, USA}}

\institution{PU}{\centerline{${}^{2}$Department of Physics, Princeton University, Princeton, NJ 08544, USA}}

\authors{Jonathan J. Heckman\worksat{\HarvardU}\footnote{e-mail: {\tt jheckman@physics.harvard.edu}} and Herman Verlinde \worksat{\PU}\footnote{e-mail: {\tt verlinde@princeton.edu}}}

\abstract{
We introduce a covariant non-commutative deformation of 3+1-dimensional conformal field theory. The deformation  introduces a short-distance scale $\ell_p$, and thus breaks scale invariance,
but preserves all  space-time isometries. The
non-commutative algebra is defined on space-times with non-zero constant
curvature, i.e.  $dS_4$ or $AdS_4$. The construction makes essential use of the representation of CFT tensor operators  as polynomials in an auxiliary polarization tensor. The polarization tensor takes active part in the non-commutative algebra, which for $dS_4$ takes the form of 
$so(5,1)$, while for $AdS_4$ it assembles into  $so(4,2)$. The structure of the non-commutative correlation functions hints that the deformed theory contains gravitational interactions and a Regge-like trajectory of higher spin excitations. }

\maketitle

\def\IJKLMN{{\mbox{\fontsize{7pt}{.5pt}$IJKLMN$}}}
\def\IJKL{{\mbox{\fontsize{7pt}{.5pt}$IJKL$}}}

\def\IJ{{\mbox{\fontsize{7pt}{.5pt}$I\nspc J$}}}
\def\JI{{\mbox{\fontsize{7pt}{.5pt}$JI$}}}
\def\JK{{\mbox{\fontsize{7pt}{.5pt}$JK$}}}
\def\KL{{\mbox{\fontsize{7pt}{.5pt}$KL$}}}
\def\IL{{\mbox{\fontsize{7pt}{.5pt}$IL$}}}
\def\JL{{\mbox{\fontsize{7pt}{.5pt}$JL$}}}
\def\IK{{\mbox{\fontsize{7pt}{.5pt}$IK$}}}
\def\II{{\mbox{\fontsize{7pt}{.5pt}$I$}}}
\def\JJ{{\mbox{\fontsize{7pt}{.5pt}$J$}}}
\def\KK{{\mbox{\fontsize{7pt}{.5pt}$K$}}}
\def\KJ{{\mbox{\fontsize{7pt}{.5pt}$KJ$}}}

\def\LL{{\mbox{\fontsize{7pt}{.5pt}$L$}}}

\def\MN{{\mbox{\fontsize{7pt}{.5pt}$MN$}}}
\def\ppartial{\raisebox{-2pt}{$\partial$}}
\def\Xbar{\, {\overline{\! X\nspc}\spc}}
\def\Sbar{\, {\overline{\! Z\nspc}\spc}}
\def\ZS{Z}
\def\cZ{{Z}}
\def\cbZ{\spc \overline{\! Z\!}\spc}

\def\bY{\spc \overline{\nspc Y \nspc} \spc} 

\enlargethispage{\baselineskip}

\setcounter{tocdepth}{2}

\renewcommand\Large{\fontsize{15}{17}\selectfont}

\renewcommand\large{\fontsize{14}{15}\selectfont}

\noindent
{\large \bf 1. Introduction \label{sec:INTRO}}
\vspace{2mm}

One of the basic ways to generalize the classical notion of space-time is non-commutative geometry  \cite{NC}. An especially attractive feature of non-commutative generalizations of quantum field theory is the natural appearance of a minimal resolution length scale. Non-commutativity may thus act as a UV regulator.  However, most known implementations of space-time
non-commutativity have the substantial drawback that they explicitly violate Lorentz invariance. For many reasons, it would therefore be of special interest to find examples of covariant non-commutative space-times,  that preserve all global symmetries of the underlying classical space-time \cite{yang,snyder,kappapoincare}.

 In this paper we propose a covariant 
non-commutative (CNC) deformation of a general 3+1-D conformal field theory (CFT) defined a homogeneous space-time with constant positive or negative curvature.  We focus on the application 
to 3+1-D de Sitter space-time $dS_4$, though the discussion can be easily generalized to
3+1-D anti-de Sitter space-time $AdS_4$. 3+1-D de Sitter space-time is described by an embedding equation of the form
\bea
\eta^{\smpc \AB}X_\AA X_\BB\is R^2 \label{embed}%
\eea
where {\footnotesize $A$, $B$} run from $0$ to $4$, $\eta_{\spc \AB}= (-,+,+,+,+)$ is the 5D Minkowski metric, and $R$ specifies
the curvature radius.  For $AdS_{4}$, we take $\eta_{\smpc \AB} = (-,+,+,+,-)$, and replace
$R^{2}\rightarrow-R^{2}$. 

\def\NC{{\mbox{\tiny NC}}}
Like any known non-commutative deformation of space-time, we will postulate that the position operators $X_\AA$ satisfy a non-trivial commutation relation of the general form
\bea
\label{cnco}
\bigl[\spc  X_{\AA},X_{\BB}\spc \bigr]  \is i \spc \hbar  \spc \ell^2  \, S_{\AB}. \label{CNC}%
\eea
Here $\ell$ is a short-distance length scale, and $S_\AB$ is a (dimensionless) anti-symmetric tensor, with $\AAn,\BBn=0,...,4$. If $S_\AB$ where some {\it fixed} anti-symmetric tensor, equation (\ref{cnco}) would break Lorentz invariance.
To obtain a covariant non-commutative deformation, the isometry group of 3+1-D de Sitter  space-time would need to act both on the coordinates $X_\AA$ and the tensor $S_\AB$,
via the infinitesimal $SO(4,1)$ generators $M_\AB$
\bea \label{momact} 
 & & \!\!\!\!\!\!\!\!\! \lbrack M_{\AB},X_{\CC}] \; = \; i\spc\hbar\spc \bigl(  \eta_{\smpc \AC}X_{\BB}-\eta_{\smpc \BC}X_{\AA}\bigr)  ,\\[3mm]
\left[  M_{\AB}, S_{\CD}\right]  \is { i}\spc  \hbar \spc \bigl(  \eta_{\smpc \AC} S_{\BD}+\eta_{\smpc \BD}S_{\AC}%
-\eta_{\smpc \AD}S_{\BC}-\eta_{\smpc \BC}S_{\AD}\bigr)  \, .
\label{momac}
 \eea
However, to justify its non-trivial transformation property, $S_\AB$ should represent an active degree of freedom similar to the space-time coordinates $X_\AA$. How can this be arranged?

Recent studies of CFT correlation functions of primary tensor operators \cite{embed}
 have made successful use of an extension of the embedding formalism \cite{dirac, ferrara}, in which the tensor operators are encoded by polynomials in an auxiliary polarization vector $P_\AA$. Tangent directions to 
 the de Sitter embedding equation (\ref{embed}) can be folded into anti-symmetric tensors 
 $S_\AB = X_{[\AA} P_{\BB]}$.  Tensor operators ${\cal O}_{\AA_1..\AA_j}(X)$ can thus be promoted into functions $ {\cal O}(X,S)$ defined on an extension of space-time $TdS_4$.
This formalism has the payoff that  CFT correlators of tensor operators 
 take the form of correlators of scalar operators  $ {\cal O}(X,S)$ on the extended space-time \cite{embed}.
Space-time isometries act on these correlators by transforming all positions and spin variables simultaneously, as in equation (\ref{momact})-(\ref{momac}).

This use of an extended space-time sheds new light on  how to obtain covariant non-commutativity. It is natural to look for possible ways to identify the anti-symmetric  tensor $S_\AB$ that encodes the spin of  tensor CFT operators with the tensor $S_\AB$ that appears in the space-time commutator algebra (\ref{cnco}). In this paper we will show that this idea can indeed be utilized to construct a covariant non-commutative deformation of a general 3+1-D CFT.

To obtain a self-consistent implementation, the spin variables $S_\AB$ also need to acquire a non-trivial commution relation.  The non-commutative version of $S_\AB$ is obtained from its commutative cousin via the replacement 
\bea
\label{smadd}
S_\AB \! & \to &  \spc S_\AB +   M_\AB
\eea
with $M_\AB$ the $so(4,1)$ generators (\ref{momact}). The quantized $S_\AB$ thus satisfy an $so(4,1)$ algebra.
Combined with equation (\ref{cnco}), the total CNC algebra extends to $so(5,1)$, the Lie algebra of the Lorentz group in 6 dimensions, under which the space-time coordinates and spin variables  transform as an anti-symmetric tensor $Z_\IJ$, defined via $Z_{5\AA} =\spc  X_\AA$ and $Z_\AB \spc = \spc \ell \spc S_\AB\, .$  

The scale of non-commutativity is set by the short distance length scale $\ell_p = \hbar\spc \ell$.   We also introduce the  dimensionless ratio 
\bea
\label{elpee}
N = R/ \ell_p .
\eea
We will assume that $\hbar$ is small and that $N$ is very large.

The interpretation of the $so(5,1)$ Lie algebra as a covariant non-commutative space-time algebra was first proposed in a short note by C.N. Yang, published in 1947 \cite{yang}, soon after Snyder's earlier work \cite{snyder}.
Somewhat surprisingly, possible concrete physical realizations of Yang's proposal have to our knowledge not
been actively investigated since.

This paper is organized as follows. In section 2 we summarize the embedding formalism for tensor operators in 3+1-D CFT and introduce the $SO(5,1)$ invariant notation. In section 3 define the covariant non-commutative space-time algebra, and write it in terms of local  Minkowski coordinates. In sections 4  and 5  we construct the star product and use this to define the covariant non-commutative deformation of the CFT correlation functions. 
In section 6 we present a spinor formulation of the covariant non-commutative space-time algebra.
We end with some concluding comments in section 7.

\bigskip
\bigskip

\def\IJ{{\spc I\nspc J}}

\def\hf{{\textstyle \frac 1 2}}
\def\Vbar{ \spc \overline{\nspc Z \nspc}\spc }
\def\mfh{\mathfrak{h}}

\newcommand*{\Scale}[2][4]{\scalebox{#1}{$#2$}}

\bigskip
\bigskip

\noindent
{\large \bf 2. Tensor Operators as Polynomials}
\vspace{3mm}

Here we briefly summarize the generalized embedding formalism for CFT correlators of tensor primary operators. A more detailed discussion can be found in \cite{embed}.

Let $X_\AA$ denote the 5D embedding coordinates of $\mathbb{R}^{4,1}$, in which 3+1-D de Sitter space is defined as the subspace (\ref{embed}).
A general spin $j$ primary operator ${\cal O}_{\AA_1\ldots \AA_j}(X)$ in a CFT
defined on 3+1-D de Sitter space-time (\ref{embed}) is symmetric, traceless and transverse 
\bea
\label{transc}
X^{\AA} {\cal O}_{\AA\smpc \AA_2\ldots \AA_j}(X) = 0, 
\quad & &  \quad 
\eta^{\spc \AB} {\cal O}_{\AA\smpc \AA_2 .. \BB ..\spc \AA_j}(X) = 0
.
\eea
The transversality constraint ensures that the tensor indices represent tangent vectors to the space-time manifold $dS_4$.
Any such tensor operator can be represented as a polynomial in an auxilary polarization vector $P_\AA$ via
\bea
{\cal O}(X,P\smpc ) = {\cal O}_{\AA_1\ldots \AA_\ell}(X) P^{\AA_1} \cdots P^{\AA_j}\, .
\eea
As long as the $P_A$ mutually commute, the polynomial ${\cal O}(X,P\smpc )$ automatically represents a symmetric tensor.  
The transversality  and tracelessness conditions (\ref{transc}) furthermore imply that the polynomials ${\cal O}(X,P\smpc )$ satisfy the differential equations
\bea
\label{trans}
\qquad \qquad X_\AA\,\frac{\partial\, }{\! \partial P_\AA\!\!}\;  {\cal O}(X,P\smpc)\is 0\qquad \quad \mbox{(transverse)} \\[3mm]
\qquad \qquad \frac{\partial^{\spc 2}\, }{\!\! \partial P^\AA\partial P_\AA\!\!\!}\;\, {\cal O}(X,P)\is  0\, \qquad \quad  \mbox{(traceless)}
\label{harm}
\eea
Traceless tensors thus correspond to harmonic polynomials in the polarization vector $P_\AA$.

The transversality condition (\ref{trans}) implies that ${\cal O}(X,P+ \alpha X) = {\cal O}(X,P)$ for any $\alpha$. Solutions to this condition are characterized by the property that the polarization variable $P_\AA$ only appears in the anti-symmetric combination \cite{embed}
\bea
\label{sxp}
S_\AB \is X_\AA P_\BB - X_\BB P_\AA .
\eea
So instead of using the vector $P_\AA$, we can choose to write the transverse tensor operators as functions ${\cal O}(X,S)$. The property that  $S_\AB$ takes the form (\ref{sxp}) is enforced by requiring that
\bea
\label{special}
\varepsilon^{\ABCDE}S_{\AB}X_{\CC} \is 0.
\eea
One readily sees that this condition implies that $S_\AB$ must have one leg in the $X_\AA$ direction. Note  that it automatically follows that $\varepsilon^{\ABCDE} S_{\AB} S_{\CD}=0$.  In terms of the variable $S_\AB$, the tracelessness condition (\ref{harm}) takes the form
\bea
\label{newharm}
\Bigl( X^{\AA}\,  \frac{\partial\  }{\! \partial S^\AB\!\!\!\!\!\!\!}\;\;\;\, \Bigr)^2  {\cal O }(X,S) = 0\, .
\eea

To build in the anomalous scale dependence of CFT correlation functions, it is often convenient to extend the embedding formalism to a 6D space-time $\mathbb{R}^{4,2}$ with (4,2) signature $(-,+,+,+,+,-)$, by adding a 
time-like coordinate $X_5$. The 6D coordinates are then restricted to lie on the null cone $X^\AA X_\AA - X_5^2 = 0,$
 on which the conformal group $SO(4,2)$ naturally acts \cite{dirac, ferrara}.
Conformal operators are  homogenous functions of dimension $-\Delta$
\bea
\label{sixscale}
{\cal O}(\lambda X, \lambda S) = \lambda^{-{\Delta}}\, {\cal O}(X,S)\,  .
\eea
The projection down to $dS_4$ amounts to gauge fixing the projective symmetry by setting $X_5 = R$. Below, we will work in the gauge fixed formulation. This is sufficient for our purpose, since  scale invariance will be broken anyhow by the non-commutative deformation.

\medskip

\underline{Summary}: We  can represent primary tensor operators  on 3+1-D de Sitter space-time as harmonic functions of the generalized coordinates $(X_\AA,S_\AB)$ subject to the constraints (\ref{embed}) and (\ref{special}).
As a simple example of how the formalism works,  the 2-point function of two spin $j$ operators with conformal dimension $\Delta$ is given by \cite{embed}
\bea
\label{spinjtop}
{\Bigl\langle {\cal O}(X_2,S_2) \, {\cal O}(X_1,S_1) \Bigr\rangle} \is {const.}\;  {\frac{ \bigl(S_1 \cdot S_2\bigr)^j\!\!\!}{\bigl(R^2 \nspc- X_1\nspc \cdot\nspc X_2\bigl)^{\Delta+j}\!\!\!\!\!\!\!\!\!\!{}^{\raisebox{6pt}{${}$}}}}.
\eea
For now, the coordinates $X_\AA$ and  spin variables $S_\AB$ seem to stand on rather different footing: the former refer to actual space-time points, while the latter are just a convenient packaging of polarization indices. The two types of variables can however be naturally unified as coordinates of an extended space-time, given by
the tangent space $TdS_4$ to 3+1-D de Sitter. We can think of the extended space-time $TdS_4$ as analogous to superspace,
used for grouping supermultiplets in supersymmetric theories into single superfields, except that now the extra polarization variables are bosonic rather than fermionic.

\def\ri{\bigr |}

\bigskip
\bigskip

\noindent
{\it SO(5,1) Symmetric Notation}
\vspace{2mm}
\def\NL{{{\nspc }_{\rm S}}}

\def\NC{{{\nspc }_{\rm P}}}

One may assemble the coordinates and spin variables into a single anti-symmetric $6\times6$ matrix  $Z_{IJ} = - Z_{JI}$, with {\small  $I,J=0,...,5$}, via 
\bea
\label{sofive}
Z_{5 \AA} \, = \, X_\AA, \quad\ & & \ \quad Z_\AB \spc = \spc  \ell \spc \smpc S_\AB .
\eea
Here $\ell$ denotes an infinitesimal expansion parameter with the dimension of length. 
To raise and lower indices, we introduce the flat metric $\eta_{\smpc IJ}$ with signature $(-,+,+,+,+,+)$. 
The embedding equation (\ref{embed}) and tranversality constraint (\ref{special}) 
can be recognized as the $\ell \to 0$ limit of the following $SO(5,1)$ invariant relations 
\bea 
\label{nspecial}  \half Z_{\IJ}\spc Z^{\IJ} \, =\,    R^2,  \qquad  & & \qquad 
 \varepsilon^{\IJKLMN} Z_{\KL}\spc  Z_{\MN} \, = \, 0\, .
\eea 
Taking $\ell \to 0$ amounts to performing a Inonu-Wigner contraction of $SO(5,1)$, which yields $ISO(4,1)$, the Poincar\'e group of the 4+1-D Minkowski space,
the embedding space of $dS_4$.   
The 8D extended space-time parametrized by $Z_{\IJ}$ then reduces  to the tangent space $TdS_4$.

The above $SO(5,1)$ symmetric notation will prove to be convenient for our purpose, but the symmetry is of course explicitly broken to the $SO(4,1)$ subgroup: interactions and correlation functions of a typical CFT do not respect the full $SO(5,1)$ or $ISO(4,1)$ symmetry. The following two basic statements remain true, however:

\medskip
\medskip

\noindent  
${}$~~\parbox{16cm}{\addtolength{\baselineskip}{1mm} 
  $\bullet$ Correlation functions of tensor operators in a 3+1-D CFT can be written as correlators
\bea
\label{correlz}
\Bigl\langle\, {\cal O}_n(Z_n)   \,...\,  {\cal O}_2(Z_2)  {\cal O}_1(Z_1) \,\Bigr\rangle 
\eea
of scalar operators defined on the 8-dimensional extended space-time $TdS_4$, parametrized by the  coordinates (\ref{sofive}),
subject to (\ref{nspecial}). 
}

\medskip
\medskip

\noindent  
${}$~~\parbox{16cm}{\addtolength{\baselineskip}{1mm} $\bullet$ Any pair of points $Z_1$ and $Z_2$ on the extended space-time are related via an 
$SO(5,1)$ rotation $\Lambda_{12}$ (or after performing the IW contraction, an $ISO(4,1)$ transformation) via
\bea
\label{ztrans}
Z_{1}^{\IJ} \is (\Lambda_{12})^\II_{\spc\KK}  \spc Z_2^{\KL}  (\Lambda_{21})^{\spc \JJ}_{\LL}\,.
\eea
Here  $(\Lambda_{12})^\II_{\spc\KK}  \spc (\Lambda_{21})^{\KK\JJ} = \eta^\IJ$. So $Z^\IJ$ transforms in the adjoint representation.}

\medskip
\medskip

\noindent
The above two statements are all that we need to proceed with our construction.

\medskip

Note that in order to obtain finite correlators from (\ref{correlz}), we need to first extract a power of $(\ell)^j$ from each spin $j$ tensor operator before taking the  Inonu-Wigner limit $\ell \to 0$. Even if $SO(5,1)$ is not a full symmetry, the $Z_\IJ$ notation sometimes still provides a convenient packaging of higher spin correlators. E.g. the 2-point functions (\ref{spinjtop}) of spin $j$ tensor operators can be be summarized into a single quasi-$SO(5,1)$ invariant formula via
\bea
\label{twozj}
\Bigl\langle\spc {\cal O}(Z_2)\, {\cal O}(Z_1)\spc  \Bigr\rangle  \is {const.} \left[\frac{1} {\bigl(R^2 - Z_1 \! \cdot \! Z_2 \bigr)^{\Delta}\!\!\!\!\!}\;\;\,\right]_{j}
\eea
where the subscript $j$ indicates the projection onto the term proportional to $(\ell)^{2j}$. This projection breaks $SO(5,1)$ to $SO(4,1)$. We will use the formula (\ref{twozj}) later on.

\medskip

For CFTs defined on $AdS_4$, the $Z_\IJ$ notation is naturally invariant under $SO(4,2)$ instead of $SO(5,1)$. Given that $SO(4,2)$ is identical to the conformal group in 3+1 dimensions, it is perhaps tempting to look for
a relation between the above  6D notation and the more conventional 6D embedding formalism for 3+1-D CFTs \cite{embed, dirac, ferrara}. However, it is important  to not confuse the two. While both 6D notations and associated symmetry groups include the space-time isometries $SO(4,1)$ (for $dS_4$) or $SO(3,2)$ (for $AdS_4$) as a subgroup, the two extended rotation groups are really distinct. Conformal transformations, including the conformal boosts, are true
symmetries  of the CFT, but do not mix coordinates and spin variables. The 6D rotations (\ref{ztrans}) that act on the $Z_\IJ$ coordinates  (\ref{sofive}), on the other hand, are not all symmetries: besides space-time isometries, equation (\ref{ztrans}) includes transformations that mix the coordinates and spin variables. The latter symmetries are explicitly broken.

\bigskip
\bigskip
\bigskip

\noindent
{\large \bf 3. Covariant  Non-Commutative Space-Time \label{sec:CNC}}
\vspace{3mm}

We now introduce the covariant non-commutative deformation of the extended space time. 
The deformation involves two distinct steps. First we postulate that the coordinates and spin variables satisfy non-trivial commutation relations.  Secondly, we replace the de Sitter space to the $SO(5,1)$ invariant embedding equations (\ref{nspecial}).
 To indicate the transition to quantized space-time, we introduce a dimensionless expansion parameter $\hbar$.
We also identify a short distance length scale $\ell_p$ related to the de Sitter radius and the scale $\ell$ via
\bea
\label{newell}
R \is N \spc {\ell_p}, \qquad \qquad \ell_p = \hbar \spc \ell.
\eea
We will assume that $\hbar$ is small and that $N$ is extremely large, so we will often work to leading order in $1/N$. We adopt the
$SO(5,1)$ invariant notation (\ref{sofive}). 

\medskip
 
The CNC deformed theory is obtained by promoting the generalized coordinates $Z_{\IJ}$ to quantum operators that satisfy  commutation relations isomorphic to the $so(5,1)$ Lie algebra
\bea
\label{zcom}
[ Z_{\IJ} , Z_ {\KL}] \is  {i \ell_p}  \spc \bigl(  \eta_{\smpc \IK} Z_{\JL}+\eta_{\smpc\JL} Z_{\IK} -\eta_{\smpc \IL} Z_{\JK}-\eta_{\smpc \JK} Z_{\IL}\bigr). \eea
In terms of the coordinates $X_\AA$ and spin variables $S_\AB$, the commutator algebra reads \cite{yang}
\bea
\label{cnce}
\bigl[\smpc  X_{\AA},X_{\BB}\spc \bigr] \spc = \spc  {i \hbar \spc \ell^2}  \spc S_{\AB}, \hspace{-1.4cm} & & \qquad\qquad\  \ \ \
\lbrack S_{\AB},X_{\CC}]\spc =\spc {i}\hbar \spc \bigl(  \eta_{\smpc \AC}X_{\BB}-\eta_{\smpc \AB}X_{\CC}\bigr)  ,\\[5mm]
\left[  S_{\AB}, S_{\CD}\right]  \is { i}\hbar \spc \bigl(  \eta_{\smpc \AC} S_{\BD}+\eta_{\smpc \BD} S_{\AC}%
-\eta_{\smpc \AD} S_{\BC}-\eta_{\smpc \BC} S_{\AD}\bigr)  \, . 
\label{cnct}
\eea
In addition to postulating the above operator algebra, we deform the embedding equation of the de Sitter space-time to its $SO(5,1)$ invariant version (\ref{nspecial}), with $\ell$ small but finite.

\medskip

As with any quantum deformation, it is important to verify the correspondence principle, that is,  that there exists a limit in which the quantum deformed theory reduces to the classical theory. In the above parametrization, we can consider two different classical limits. We can

\noindent
(i) set the length scale $\ell = 0$, while keeping the non-commutativity parameter $\hbar$ finite, or
 
 \noindent
 (ii) 
turn off the non-commutativity by setting $\hbar =0$, while keeping the length scale $\ell$ finite.

\noindent
It is clear that if we take both limits at the same time, we get back the undeformed CFT. But what happens if we
take only one of the two limits? We claim that in both cases, we recover the undeformed CFT by performing a simple similarity transformation.

Taking limit (i) produces a scale invariant theory with commutative coordinates $X_\AA$, but with non-commutative spin variables $S_\AB$. The correlation functions are directly obtained from the commutative correlators by performing the replacement (\ref{smadd}), with $M_\AB$ the vector field that 
implements the $SO(4,1)$ rotations (\ref{momact})-(\ref{momac}). This redefinition does not amount to a true deformation of the CFT, but rather to a natural extension of the embedding formalism, in which the descendant
operators (obtained by acting with the $M_\AB$ generators on primary operators) are mixed 
in with tensor primary operators. In other words, the geometric meaning of $S_\AB$ has been deformed, rather than the CFT.

Taking limit (ii) produces a commutative theory, defined on the deformed de Sitter space-time (\ref{nspecial}). 
For given value of $S_\AB$, equation (\ref{nspecial}) amounts to a rescaling 
 of the de Sitter radius
\bea
\label{rscale}
R  \,  \to \, \lambda_S \spc R \qquad & {\rm with} & \qquad \lambda_S = \left(1 - \frac{\spc S_\AB S^\AB}{2\hbar^2 N^2} \right)^{\! 1/2}\, .
\eea 
Thanks to the conformal invariance of the original CFT, this geometric deformation  can be absorbed into a conformal transformation of local operators ({\it c.f.} equations (\ref{sixscale}) and (\ref{spinjtop}))
\bea
\label{oscale}
{\cal O}(X,S)\; \to\; \lambda_S^{\Delta+j}\, {\cal O}\bigl(\lambda_S X ,S\bigr).
\eea
CFT correlation functions are invariant under the combined transformation (\ref{rscale}) and (\ref{oscale}). We conclude that taking limit (ii) again produces the commutative CFT.

The actual size of the CNC deformation is set by $\ell_p = R/N$. 
The other constants $\hbar$  and the scale $\ell$ are just useful formal parameters. Indeed,
from the commutator (\ref{cnce}) and Heisenberg, we learn that the coordinates $X_\AA$ become fuzzy at the scale
\bea
\la\, \bigl(\Delta X_\AA\bigr)^2 \, \ra \!&\! \gsim\! & \!  \ell_p^2 \, \sqrt{C_2}, \qquad {\rm with} \qquad C_2 = \frac 1 {2\hbar^2}\spc S_\AB S^\AB. 
\eea 
We can identify $C_2$ with the second Casimir of the $SO(4,1)$ space-time isometry group.

In the next sections, we will  translate the CNC algebra of the generalized coordinated $(X,S)$ into an explicit and well-defined deformation of the CFT correlation functions.

\bigskip
\bigskip

\noindent
{\it Flat Space Limit \label{sec:FLAT}}
\vspace{-1mm}

It is instructive to write the CNC deformation in a local Minkowski space-time region, say, where $X_4 \simeq R$ and $X_\mu \ll R$ for $\mu =0, .., 3$. 
We thus isolate the $X_4$ coordinate and use $SO(3,1)$ notation.  To write the commutation relations we identify $S_\AB$ with the  $SO(4,1)$ symmetry generators $M_\AB $,
into 6 Lorentz generators/angular momenta $J_{\mu\nu} =  M_{\mu\nu}$ and 4 translations/momenta $P_\mu =  \frac 1 R M_{4\mu }$. 
 The special relations (\ref{nspecial}) imply that angular momentum is purely orbital angular momentum  $J_{\mu\nu} = P_{[\mu} X_{\nu]}$.  
 We also introduce the notation $\mfh = \frac \hbar R X_4$. Indeed,  $\mfh$ will play the role of Planck's constant (not to be confused with our $\hbar$ parameter).
 
The positions and momenta satisfy the following covariant commutation relations
\bea
[  X_{\mu},P_{\nu}]  =i{\mathfrak{h}}\hspace{1pt}\eta_{\smpc \mu\nu
},\qquad\quad\left[  P_{\mu},\hspace{1pt}P_{\nu}\hspace{1pt}\right]\! \is\! \frac
{i \hbar }{R^{2}\! }\, J_{\mu\nu},\quad\qquad\left[  X_{\mu},X_{\nu}\right]
= i\hbar \spc \ell^{2}\hspace{1pt}J_{\mu\nu}.
\eea
The first and second relation look like the standard commutation relations for coordinates and momenta in
a local patch of de Sitter space. The third relation is the covariant non-commutative deformation.
In the above notation, it appears that translation invariance is broken, because the Lorentz generators
$J_{\mu\nu}$ act relative to a preferred origin in space-time. Translation symmetry is preserved, however,
because (i) it acts via $X_\mu \to X_\mu + \mfh\spc a_\mu$, and (ii)  the effective Planck constant $\mfh$ is an operator with non-trivial commutation relations   \cite{yang} 
\bea
\left[ \spc {\mathfrak{h}}\spc , X_{\mu}\right]  =i\hbar\spc \ell^{2}\spc P_{\mu}%
,\qquad\ \ \ \ \ \ \left[ \spc {\mathfrak{h}}\spc ,P_{\mu}\right]
=-\frac {i\hbar} {R^{2}}\spc X_{\mu},\text{ \ \ \ \ \ \ \ \ \ }\left[ \spc {\mathfrak{h}%
}\hspace{1pt},J_{\mu\nu}\right]  =0.
\eea
The magnitude of ${\mathfrak{h}}$ is determined by the Casimir relation (\ref{nspecial})
\bea
{\mathfrak{h}}^{2}\is \hbar^2 \Bigl(1-\frac{1}{R^{2}}X_{\mu}X^{\mu}-\ell^{2}P_{\mu}P^{\mu}\Bigr) - \frac{1}{2N^2}J_{\mu\nu}J^{\mu\nu}. \label{hmag}%
\eea
Since, without loss of generality, we can assume that $X^\mu \ll R$, we see that  ${\mathfrak{h}} \simeq \hbar$  as long as  the mass squared
is small compared to $1/\ell^2$.  We further note that the above algebraic relations enjoy an intriguing T-duality
symmetry under the interchange of coordinates and momenta $
\frac 1 R X_{\mu} \leftrightarrow  \ell P_{\mu}$ combined  with the reflection ${\mathfrak{h}}
\leftrightarrow-{\mathfrak{h.}}$


\bigskip
\bigskip
\bigskip

\noindent
{\large \bf 4. Star Product \label{sec:STAR}}
\vspace{2mm}

We will now construct the star product between functions on the non-commutative extended space-time. 
In its most basic form,  the star product deforms the product of two functions $G(Z)$ and $F(Z)$ evaluated at the same point $Z$ on the extended space-time. Since in our case the commutator algebra between the coordinates $Z$ takes the form of a Lie algebra, this star product is well understood and directly related to the CBH formula \cite{cbh}.

  The abstract definition is as follows. We can make a unique mapping between commutative functions $F(Z)$
and symmetrized functions $F(\hat{Z})$ of the non-commutative coordinates $\hat{Z}_\IJ$. (Here we temporarily decorate the operator valued coordinates with a hat.)
The product $G(\hat{Z}) F(\hat{Z})$ of two symmetrized non-commutative functions is not a symmetrized function in the non-commutative coordinates $\hat{Z}$. But it can always be reorderered such that it becomes symmetrized. The extra terms produced by the re-ordering process are encoded in the star product $G(Z) \star F(Z)$. 
This prescription thus identifies $(G \star F)(\hat{Z}) = (G(\hat{Z}) F(\hat{Z}))_{\rm sym}$. This rule associates a unique star product to any Lie algebra.

The resulting star product takes the following general form \cite{cbh}
\bea
\label{dstar}
G(Z)\star F(Z) \is \exp\Bigl\{\hbar \spc Z^\IJ D_\IJ \bigl(\spc\mbox{\scriptsize$\overleftarrow{\mbox{\normalsize \nspc $\partial$\spc}}$}_{\! Z} ,\mbox{\scriptsize$\overrightarrow{\mbox{\normalsize $\nspc \partial\spc$}}$}_{\! Z} \bigr)\Bigr\} \, G(Z) \spc F(Z)
\eea
where $\mbox{\scriptsize$\overleftarrow{\mbox{\normalsize \nspc $\partial$\spc}}$}_{\! Z}$ acts on $G(Z)$ and $\mbox{\scriptsize$\overrightarrow{\mbox{\normalsize $\nspc \partial$\spc}}$}_{\! Z}$ on $F(Z)$. An explicit, albeit formal, definition of the symbol $D_{IJ}(a,b)$ in terms of the CBH formula is given in \cite{cbh} and in the  Appendix. 
The CBH star product (\ref{dstar}) has an obvious generalization to an  $n$-fold star product 
$F_n(Z) \star \ldots \star F_2(Z) \star F_1(Z)$ between $n$ functions of the same variable $Z$. This $n$-fold star product is associative.

For our application, however, we will need the star product $G(Z_2)\spc \star \spc F(Z_1)$ between functions evaluated at two {\it different} locations on the non-commutative extended space-time. This may seem like a completely new concept, since a natural first guess would be to treat the two locations $Z_1$ and $Z_2$ as two independent mutually commuting operators. However, this would lead to a trivial star product. In our case, we are helped  by the fact that any two points $Z_1$ and $Z_2$ on the commutative space can be related to each other via an $SO(5,1)$ rotation. This means that we can view the non-commutative points $Z_1$ and $Z_2$ as two {\it quantum} coordinates related via a {\it classical} $SO(5,1)$ rotation $\Lambda_{12}$, as in equation (\ref{ztrans}). 

Let us pick some  fixed but otherwise arbitrary base point $Z$. We can then factorize the $SO(5,1)$ rotation $\Lambda_{12}$ that relates $Z_1$ to $Z_2$ into a product 
\bea
\Lambda_{12}= \Lambda_1 \Lambda_2^{-1}, \quad & &  \quad 
Z_1 = \Lambda_{1}\!\cdot \! Z\spc , \qquad \qquad Z_2 =  \Lambda_{2}\!\cdot \! Z\spc ,
\eea
where $\Lambda_1$ and $\Lambda_2$ are the $SO(5,1)$ rotations that relate each corresponding point to the base point $Z$. Here $\Lambda_{1}\!\cdot \! Z$ is short-hand for the action of $\Lambda_1$ on $Z$, as defined in equation (\ref{ztrans}).

We can now promote the base point $Z$ to a non-commutative coordinate, while leaving the $SO(5,1)$ rotations 
$\Lambda_1$ and $\Lambda_2$ as classical quantities, and write the star product
\bea
\label{onestar}
G(Z_2) \star F(Z_1) \is G\bigl(\Lambda_2\! \cdot \! Z\bigr) \star F\bigl(\Lambda_1 \! \cdot \! Z\bigr) .
\eea
Here the $\star$-symbol is defined as in equation (\ref{dstar}), with $G \circ \Lambda_2$ and $F \circ \Lambda_1$ treated as functions of the base point $Z$.
Note that this definition treats the points $Z_1$ and $Z_2$ symmetrically.

\def\leta{\Lambda}

The star product (\ref{onestar}) seems to depend on the choice of base point $Z$. 
It is easy to see, however, that this $Z$ dependence is spurious.
A direct way to make this explicit is by computing the commutator
$\bigl[\spc Z_1,\spc Z_2 \spc \bigr]  = Z_1 \star Z_2 - Z_2 \star Z_1$ between the two different non-commutative locations.
Using the definition (\ref{onestar}), we find that
\bea
\label{zzcom}
\bigl[\spc Z_1^\IJ,\spc Z^\KL_2 \spc \bigr] \is \imath \ell_p\spc Z_{12}^{\IJKL} 
\eea
\vspace{-1.05cm}
with
\bea
\label{zijkl}
& Z_{12}^{\IJKL}  \, = \, \leta_{\spc 12}^{\IK} \, Z^{\JL}_{12}\spc +\spc\leta_{\spc 12}^{\JL}\, Z^{\IK}_{12} \spc -\spc \leta_{\spc 12}^{\IL} \, Z_{12}^{\JK}\spc -\spc \leta_{\spc 12}^{\JK} \, Z_{12}^{\IL}, &\\[5mm]
& 
Z_{12}^\IJ \, = \, (\Lambda_{1})^{\II}_{\spc \KK} Z^{\KK\LL}(\Lambda_2)_\LL^{\spc\JJ},
&
\eea
where $\Lambda_{12} = \Lambda_1 \Lambda_2^{-1}$ is the $SO(5,1)$ rotation that relates the two positions. The algebra  (\ref{zzcom}) reduces to the standard $so(5,1)$ Lie algebra (\ref{zzcom}) in the limit $\Lambda_1 = \Lambda_2$, i.e. if $Z_1$ approaches $Z_2$.  Note further that the dependence on the base point $Z$ has dropped out: both $\Lambda_{\spc 12}^\IJ$ and $Z_{12}^\IJ$ are invariant under $SO(5,1)$ rotations acting on the base point $Z$. Hence the algebra (\ref{zzcom}) is fully covariant.

The definition (\ref{onestar}) directly generalizes to an associative $n$-fold star product between $n$-functions evaluated at $n$ different points $Z_k$, each given by a classical $SO(5,1)$ rotation $\Lambda_k$ applied to the base point $Z$. 
Since we are assuming that the deformation parameters $\ell_p = \hbar\spc \ell$ is very small, we will mostly focus on the 
first order non-commutative corrections linear in $\ell_p$. To write this linearized expression in a somewhat more explicit form, we use the obvious generalization of the formula (\ref{zzcom}) to any pair of points $Z_i$ and $Z_j$. From this we immediately derive that
\bea
\label{nstar}
F_n(Z_n) \star \cdots  \star F_1(Z_1) \is F_n(\Lambda_n \!\cdot\! Z) \star \cdots  \star F_1(\Lambda_1\!\cdot\! Z) \nonumber \\[-1mm]
\\[-1mm]
\is \left(\, 1 \; + \; \frac {\imath\ell_p}{2}  \sum_{i<j}  \, Z_{ij}^{\IJKL}\spc \frac{\! \partial \ }{ \partial Z_i^\IJ\!\!\!\!}\;\;\; \frac{\partial \  }{ \! \partial Z_j^{\KL}\!\!\!\!\!\!}\;\;\; + \ldots\right)  F_n(Z_n) \cdots  F_1(Z_1)\nonumber\ \ 
\eea
The non-commutativity of the product is reflected by the fact that the right-hand side depends on the ordering of the $n$-tuple of coordinates $Z_1, Z_2, \ldots,  Z_n$.  

\bigskip
\bigskip
\bigskip

\def\ppsi{\psi}

\noindent
{\large \bf 5. Non-Commutative CFT Correlators}
\vspace{2mm}

We now define the CFT $n$-point functions on the covariant non-commutative space-time
as the non-commutative deformation of the CFT correlators on commutative space-time.
Concretely,  we can first decompose the commutative correlation function (\ref{correlz})
as a sum of products of ordinary functions of the positions $Z_k$,  by inserting a complete set of CFT states $\sum_\psi | \spc \ppsi \spc \rangle \langle \spc \ppsi \spc |$ in each intermediate channel. The result is a sum of factorized terms of the form
\bea
\label{mprod}
\la 0 \ri \, {\cal O}_n(Z_n)\spc \li \ppsi_n\ra  
\,\ldots\, 
 \la \smpc \ppsi_2 \smpc \ri {\cal O}_2(Z_2)\li \ppsi_1 \ra \la  \ppsi_1 \ri{\cal O}_1(Z_1)  \li  0\ra
\eea
Using the definition of the star product in the previous section, we can now directly write the definition of the  CFT $n$-point functions on the covariant non-commutative space-time by replacing 
each  factorized  product (\ref{mprod}) by the corresponding $n$-fold star product (\ref{nstar}):
\bea
\label{mprods}
\la 0 \ri \, {\cal O}_n(Z_n)\spc \li \ppsi_n\ra \star   
\,\ldots\, 
 \star \la \smpc \ppsi_2 \smpc \ri {\cal O}_2(Z_2)\li \ppsi_1 \ra  \star \la  \ppsi_1 \ri{\cal O}_1(Z_1)  \li  0\ra
\eea

Adopting this natural prescription, we obtain the following formula for the correlation functions of the CFT on covariant non-commutative space-time, expanded to linear order in the deformation parameter $\ell_p = \hbar\spc \ell$
\bea
\Bigl\langle\, {\cal O}_n(Z_n)     \,\ldots\, {\cal O}_2(Z_2) \,  {\cal O}_1(Z_1) \,\Bigr\rangle_{\rm CNC}  \, = \, 
\Bigl\langle\, {\cal O}_n(Z_n) \star \,\ldots\,\star {\cal O}_2(Z_2)  \star {\cal O}_1(Z_1) \,\Bigr\rangle_{\rm CFT}\qquad \nonumber
\\[-1mm]
\label{nccorr}
\\[-1mm]
\qquad\qquad = \, \left(1 \, + \, \frac {\imath \hbar \spc \ell}{2}  \sum_{i<j}  \, Z_{ij}^{\IJKL}\spc \frac{\! \partial \ }{ \partial Z_i^\IJ\!\!\!\!}\;\;\; \frac{\partial \  }{ \! \partial Z_j^{\KL}\!\!\!\!\!\!}\;\;\, + \ldots \right) 
\Bigl\langle\, {\cal O}_n(Z_n)    \,\ldots\,  {\cal O}_2(Z_2)  {\cal O}_1(Z_1) \,\Bigr\rangle_{\rm CFT}. \nonumber
\eea
The higher order terms can in principle be computed by expanding out each CBH star product, using equations 
(\ref{dstar})-(\ref{onestar}) and the definition of $D_\IJ(a,b)$ given in the Appendix.

The CNC deformation (\ref{nccorr})  breaks conformal invariance but is otherwise invariant under all space-time isometries. 
The right-hand side depends on the ordering of the $n$-tuple of coordinates $Z_1,Z_2, \ldots, Z_n$, or equivalently, on the  ordering of  CFT operators ${\cal O}_i(Z_i)$. It is natural to identify the non-commutative operator ordering with the time-ordering of the undeformed CFT. Note, however, that the deformed correlation functions are no longer fully crossing symmetric. 
Only the duality relations that respect the ordering of the $n$-tuple $Z_1, Z_2, \ldots, Z_n$ are preserved.  In this respect, the CNC theory is similar to open string theory. 

In abstract terms, the commutative associative operator product algebra of the original CFT is deformed into a non-commutative associative operator product algebra of the CNC theory. 
This structure is  expected from correlation functions on a non-commutative space-time. We will not try to further formalize this operator algebraic perspective in this paper.

\bigskip
\bigskip

\noindent
{\it Some Simple Examples}

To get a bit more physical insight into the nature of the CNC deformation, let us look at two simple examples: the  2-point functions of spin $j$ tensor operators (\ref{twozj}) and the $n$ point function of $n$ scalar operators.
In both cases we focus on the first order CNC correction. From now on we will use units such that de Sitter radius $R=1$, so that $\ell_p = \hbar\spc \ell = 1/N$.

The first order CNC deformation of the 2-point functions of arbitrary spinning primaries is most easily found by inserting the (quasi) $SO(5,1)$ invariant expression (\ref{twozj}) into the general formula (\ref{nccorr}), and expanding the result
\bea
\label{twocnc}
\Bigl\langle\spc {\cal O}(Z_2) \, {\cal O}(Z_1)\spc  \Bigr\rangle_{\rm CNC} \is 
\frac{1} {\bigl(1 - Z_2 \! \cdot \! Z_1 \bigr)^{\Delta}\!\!\!\!\!}\; \;\, +\, \imath\hbar\spc \ell \spc \frac{\spc \Delta(\Delta\nspc +\nspc 1)}{2} \;\, \frac{ Z_2 \! \cdot \nspc Z_{12}\! \cdot \! Z_1}{\!\! \bigr(1 - Z_1 \! \cdot \nspc Z_2 \bigr)^{\Delta+ 2}\!\!\!\!\!\!\!\!\!\!}\;\; \; \;\;\;+\, \ldots\, 
\eea
in powers of $\ell^2$. 
Here most of the $\ell$ dependence is hidden in the definition (\ref{sofive}) of $Z_\AB$ in terms of spin variable $S_\AB$.
As noted in equation (\ref{twozj}), the term proportional to $\ell^{2j}$ in the first term of  equation (\ref{twocnc}) gives the undeformend 2-point function of the spin $j$ primaries. The corresponding linearized CNC deformation is found by extracting the term proportional to $\hbar\spc \ell^{2j+2}$
from the correction term on the right-hand side of (\ref{twocnc}).

Specializing to the case $j=0$, we are instructed to look for the leading order term  in the $\ell$ expansion. A straightforward calculation gives
\bea
\qquad Z_2 \! \cdot \nspc Z_{12}\! \cdot \! Z_1  \is \ell X_2 \! \cdot \! S_{12} \! \cdot \! X_1 \; \; + \; \ldots
\eea
with
\vspace{-1.1cm}
\bea
\label{sonetwo}
S_{12}^\AB \is \half\bigl(S_1^{\AC}(\Lambda_{12})^{\; \BB}_{\CC} + (\Lambda_{12})^{\AA}_{\, \CC}\, S_2^{\CC\BB} \bigr) ,
\eea
where $\Lambda_{12}$ denotes the $SO(4,1)$ rotation that relates the space-time positions $X_1$ and $X_2$. 
This first order correction turns out to vanish, however, due to the anti-symmetry of the $S_\AB$ tensor.
The leading  correction to the scalar  2-point function thus appears at second order in the $\hbar$ expansion. 
 We will compute this correction in the next subsection, where
we will see that, due to the fact that the commutator (\ref{cnce}) between $S_\AB$ and $X_\AA$ is not suppressed by  $\ell$,  this second order CNC correction already appears at order  $\hbar^2 \ell^2$. 

The higher $n$ point functions of scalar operators do receive a non-zero first order correction. We find that it takes the following form
\bea
\label{ncorr}
\Bigl\langle\, {\cal O}_n(Z_n)\ldots    {\cal O}_2(Z_2) \,  {\cal O}_1(Z_1) \,\Bigr\rangle_{\rm CNC}  \, = \, & & \nonumber\\[-1.5mm]\\[-1.5mm]
\qquad \qquad \qquad
\Bigl(1 \, + \, \frac {\imath \hbar \ell^2}{2}  \sum_{i<j}   \, S_{ij}^{\AB}\spc \frac{\partial \, }{ \partial X_i^\AA\!\!\!\!}\;\;\; \frac{\partial \,  }{ \! \partial X_j^{\BB}\!\!\!\!\!\!}\!\!  & & +  \, \ldots \,  \Bigr) 
\Bigl\langle\, {\cal O}_n(X_n)  \ldots  {\cal O}_2(X_2) \,  {\cal O}_1(X_1) \,\Bigr\rangle_{\rm CFT} , \qquad \nonumber\\[-2mm]\nonumber
\eea
with $S_{ij}^\AB$ defined as in equation (\ref{sonetwo}). 
The undeformed $n$-point function is independent of the $S_\AB$ tensors, which reflects the spin zero property of the scalar primary operators. The appearance of the polarization tensors in the first order correction thus indicates that the CNC deformation produces new states or components with non zero spin.  We will comment on the physical interpretation of this result in the concluding section.

\bigskip
\bigskip

\noindent
{\it Second Order CNC Correction}
\vspace{-.2mm}

In the previous section, we saw that the linearized non-commutative correction to the 2-point function of scalar operators vanishes. Here we will consider the second order correction. It arises from the next order term in the CBH star product (\ref{dstar}), applied to a function with two arguments $X_1$ and $X_2$. The relevant terms are given by the double commutators in the CBH expansion  (\ref{dexp}). Since the scalar 2-point function are independent of the polarization tensors $S_\AB$, we only need to consider double commutators between the actual space-time coordinates $X_\AA$.

\smallskip

Covariant non-commutative space-time is characterized by the  basic double commutator
\bea
\label{double}
\bigl[ [ X_\AA, X_\BB], X_\CC\bigr] \is -\hbar^2 \ell^2 \spc \bigl( \eta_\AC X_\BB - \eta_\BC X_\AA\bigr),
\eea
which expresses the fact that the single commutator between $X_\AA$ and $X_\BB$ is proportional to the space-time isometry generator $M_\AB$.\footnote{This double commutator relation hints that the CNC deformation may perhaps be related 
to the presence of a 3-form potential.}

\smallskip

We will now use this formula for the double commutator to compute the leading CNC correction to the scalar CFT 2-point function
\bea
\Bigl\langle\nspc {\cal O}(X_2)\, {\cal O}(X_1)\nspc \Bigr\rangle_{\rm \! CFT} \! \is \, \frac{1} { \!\bigr( 1 \nspc -\nspc  X_1 \! \cdot \! X_2 \bigr)^{\Delta}\!\!\!\!}\;\;\;\;
\eea
To simplify our computation somewhat, we will assume that the two points $X_1$ and $X_2$ are close to each other,
so that we can drop all terms that are suppressed by a relative factor of $(1\nspc -\nspc X_1\!\cdot \! X_2) \ll 1$.
In most places, we can thus set the $SO(4,1)$ rotation $\Lambda_{12}$ that relates $X_1$ and $X_2$ equal to unity.

In this regime, we find that the second order contribution to the CBH star product ${\cal O}(X_1) \star {\cal O}(X_2)$
is given by ({\it c.f.} equations (\ref{double}) and (\ref{dexp}))
\bea
\frac {\hbar^2 \ell^2} {12} \Bigl[(X_1\!\nspc\cdot\nspc \partial_2)\spc \square_1 + (X_2\!\nspc \cdot\nspc \partial_1)\spc \square_2 - \bigl(X_1\!\nspc \cdot\nspc \partial_1 + X_2\!\nspc \cdot\nspc \partial_2\bigr) \spc (\partial_1\!\nspc\cdot\nspc \partial_2)\Bigr]\;
{\cal O}(X_2)\, {\cal O}(X_1)\spc  
\eea
with $\square = \eta^\AB \partial_\AA \partial_\BB$. A  straightforward computation then gives the following result for the 
scalar 2-point function in the CNC theory
\bea
\label{twopcnc}
\Bigl\langle\nspc {\cal O}(X_2)\, {\cal O}(X_1)\nspc \Bigr\rangle_{\rm \! CNC} \! \is  \frac{1} { \bigr( 1 \nspc -\nspc  X_1 \! \cdot \! X_2 \bigr)^{\Delta}\!\!\!\!}\;\;\;\;
\, -\; \,
\frac{{\hbar^2 \ell^2} \Delta(\Delta\nspc +\nspc 1)} { 2\spc \bigr( 1 \nspc -\nspc  X_1 \! \cdot \! X_2 \bigr)^{\Delta+2}\!\!\!\!\!\!\!\!\!} \;\;\;\;\;\; + \, \ldots 
\eea
We will comment on the possible physical significance of this result in the concluding section.

\bigskip
\bigskip
\bigskip
\noindent
{\large \bf 6. Spinor Realization \label{sec:SPINOR}}
\vspace{2mm}

\def\cX{Y}
\def\bX{\bY}

\def\wZ{W}
\def\tZ{\widetilde{Z}}
\def\bZ{\, \overline{\! Z}}
\def\bW{\, \overline{\! W\!}\,}

\def\zz{\mbox{\large $z$}}
\def\ww{\mbox{\large $w$}}

\def\cG{{\cal G}}

In this section we will describe a natural spinor realization of covariant non-commutative space time. Recall that the CNC space-time, defined by equations (\ref{sofive}), (\ref{nspecial}), represents and 8-dimensional
manifold with a symplectic form specified by equation (\ref{zcom}).
 
Introduce two independent complex four component spinors $\zz^i_\alpha$  with $i = 1,2$, $\alpha = 1,.,4$ and
 $4\times 4$ chiral gamma matrices $\Gamma_{\! \II}$ of $SO(5,1)$, satisfying the 5+1-dimensional Clifford algebra $\{ \Gamma_{\! \II}, \Gamma_{\! \JJ}\} = 2 \eta_{\smpc \IJ}$.
We define the Dirac conjugate spinors $\overline{\zz}{}^\alpha_i$ via
$\overline{\zz}_i  =  \zz_i^{\dag} \, \Gamma_{0}. $
We postulate the canonical commutator algebra\footnote{We apologize for using the same notation $i$ for the $U(2)$ index as for labeling the different locations in $n$ point functions. Hopefully this double use of notation will not lead to confusion.}
\bea
\label{zwcoms}
\bigl[ \smpc \overline{\zz}{}^{\spc i}_{\alpha}, {\zz}{}^{\spc \beta}_j\smpc \bigr]  \is \spc\imath  \hbar \spc \delta_{\alpha}^{\,\, \beta} \delta^i_{\; j} .
\eea
The spinor doublet has eight {\it complex} dimensions, thus twice the required number. 

The reduction to eight real dimensions
will be achieved by dividing out the  $U(2)$ symmetry group, which acts on the $i$ index of the $\zz^i_\alpha$ spinors, by means of a symplectic quotient. We set the four generators
 equal to a prescribed value via the constraints
\bea
\label{gcons}
\cG_i{}^j \, = \,
 \zz^\alpha_i\spc \overline{\zz}_\alpha^j - 2\spc N\spc \delta_i^{\; j} \is 0.
\eea
Here we included a constant $2N$ in the definition of the diagonal generator.
We require that all physical quantities commute with these four constraints.
Since taking the symplectic quotient  reduces the dimensionality by
twice the dimension of  the symmetry group, we end up with an 8-dimensional phase space, which
as we will now argue, coincides with the covariant non-commutative space-time with $R/\ell_p = N$.

Introduce the $4\times 4$ matrix coordinates $Z_\alpha{}^\beta$ as the $U(2)$ invariant spinor bilinears
\bea
\label{zzzdef}
Z_\alpha{}^\beta = \overline{\zz}_\alpha^i \zz_i^\beta\spc - N \delta_\alpha{}^\beta
\eea
Here we subtracted the diagonal, so that the $Z_\alpha{}^\beta$ are traceless.
One easily verifies that the $Z_\alpha{}^\beta$ coordinates commute with the $U(2)$ constraints $\bigl[\spc \cG_i{}^j ,\spc 
Z_\alpha{}^\beta\spc \bigr] = 0$,
and moreover, that on the constraint subspace ${\cal G}_i{}^j=0$,  satisfy the property
\bea
\label{zsq}
Z_\alpha{}^\gamma Z_\gamma{}^\beta \is  \, N^2 \delta_\alpha{}^\beta\, .
\eea
Finally, the commutator algebra of the $Z_\alpha{}^\beta$ coordinates takes the form
\bea
\label{zalg}
\bigl[\spc Z_\alpha{}^\beta,\spc Z_\gamma{}^\delta\spc \bigr] \is \imath\hbar\bigl( \delta_\alpha{}^\delta\spc Z_\gamma{}^\beta -   \delta_\gamma{}^\beta\spc Z_\alpha{}^\delta\bigr).
\eea
This can be recognized as the Lie algebra of $SU^*(4) \simeq SO(5,1)$.

The coordinates $Z_\IJ$ on the non-commutative extended space time are now simply obtained by taking the trace with the 
$\Gamma$ matrix bilinears 
\bea
Z_\IJ \is  \ell \tr\bigl( Z\, \Gamma_{\!\nspc \IJ} \bigr) \, = \, \ell\,\spc \overline{\zz}_\alpha^i\, (\Gamma_{\!\nspc \IJ})^\alpha{}_\beta\, {\zz}_i^\beta,
\eea
where $\Gamma_{\IJ} = \frac 1 2 [\spc \Gamma_{\! \II},\Gamma_{\! \JJ}\smpc ]$ and $\ell$ is the short-distance length scale. One directly verifies that the relation (\ref{zsq}) is equivalent to the special properties (\ref{nspecial}). 

The above spinor representation of the space-time coordinates looks quite similar to the twistor parametrization of the 6D embedding space, used in recent studies of CFT correlators \cite{shadows}. However, as emphasized earlier, the extended $SO(5,1)$ symmetry 
has no obvious relation with the conformal group $SO(4,2)$, except that both share the de Sitter isometry group $SO(3,1)$ as a common subgroup. The $z_i^\alpha$ coordinates should therefore not be confused with twistor coordinates ({\it c.f.} \cite{penrose} and \cite{us}).

Since the CFT primary operators are defined on the extended space-time, they lift to $U(2)$ invariant functions of $z$. Hence physical operators ${\cal O}(\zz)$ commute with the constraints: 
\bea
\bigl[\spc 
\cG_i{}^j ,\spc {\cal O}(\zz)\bigr] = 0.
\eea
The benefit of the spinor parametrization is that the symplectic form, that defines the CNC deformation, now has an even simpler form (\ref{zwcoms}). In particular, we can write a more explicit expression for the star product between operators ${\cal O}(z)$ and ${\cal O}(w)$ evaluated at two different points. As before, we introduce the $SU^*(4)$ rotation 
that relates the two locations
\bea
\label{rotz}
\ww_i^\alpha \spc = \spc \Lambda^\alpha_{\, \beta}\, \zz{}_i^{\spc \beta} , \quad & & \quad \overline{\zz}^{\spc i}_\alpha \spc = \spc \Lambda_\alpha^{\; \beta}\, \overline{\ww}{}^{\spc i}_\beta\, .
\eea
Since the constraints  (\ref{gcons})  are $SU^*(4)$ invariant, this rotation acts within the constraint manifold.
From (\ref{zwcoms})-(\ref{rotz}), we read off that 
$[\overline{\zz}^{\spc i}_\alpha, \spc \ww_j^\beta ]   = \imath  \hbar \Lambda_\alpha^{\; \beta}\,  \delta^{\spc i}_{\, j}$ and $[{\ww}_i^\alpha, \overline{\zz}{}_\beta^{\spc j} ] = - \imath  \hbar \Lambda_{\, \alpha}^{\beta} \, \delta_i^{\; j} .$
The star product between CFT operators defined at different points is thus given by
\bea \label{zstar} {\cal O}(\zz) \star
 {\cal O}(\ww) \is {\cal O}(\zz) \star
 {\cal O}(\Lambda\!\cdot\! \zz)\nonumber\\[-1.5mm]\\[-1.5mm]
 \is \exp \left\{ \, \imath \hbar \spc \Lambda_\alpha^{\; \beta}\, \frac{\partial\ }{\! \partial {\zz^{\spc \alpha}_{ i}}\!\!} \;\;
\frac{\partial\ \, }{\! \partial{\,\overline{\ww}_{\beta}^{\, i}}\!\!} \; - \,  \imath \hbar \Lambda_{\; \beta}^{\alpha} \,   \frac{\partial\ }{\!\partial{\overline{\zz}_{\alpha}^{\, i}}\!}\;\, \frac{\partial\ \,}{\! \partial {\ww^{\spc \beta}_{i}}}  \,
\right\}
\, {\cal O}(\zz) \,
{\cal O}(\ww), \nonumber
\eea
where we are instructed to treat $\Lambda$  as independent of $\zz$ and $\ww$. 
The formula (\ref{zstar}) easily generalizes to $n$ point functions. The fact that we can write the star product in more explicit form, suggests that the spinor formulation could be a powerful tool in deriving exact all order expressions for the CNC deformed 2 and 3-point functions.

\bigskip
\bigskip
\bigskip

\noindent
{\large \bf 7. Conclusion}

In this paper we have introduced a covariant non-commutative deformation of 3+1-D space-time  with  constant
curvature.  We have defined a general  CFT
on the non-commutative space-time, by using of the representation of arbitrary spin CFT operators as polynomials in an auxiliary polarization tensor. In fact, most of our construction (except for the spinor description) immediately generalizes to conformal field theories in arbitrary number of space-time dimensions.
In this concluding section, we make some further comments on the posible physical significance of the CNC deformation. We list the comments as questions.

\bigskip

\noindent
{\it  Minimal length?}

The non-commutative space-time algebra for $dS_4$ assembles into  $so(5,1)$, whereas for $AdS_4$ it   becomes an $so(4,2)$ algebra. The difference between the two is related to the fact that the constant time slice of 
$dS_4$ is a compact 3 sphere $S^3$, whereas for $AdS_4$ it is a non-compact hyperbolic space. In particular, the algebra of the four spatial embedding coordinates $X_a$ (in which the $S^3$ is embedded via $\eta^{ab} X_a X_b = 1$), together with the $SO(4)$ isometries of $S^3$, generate the Lie algebra of the compact $SO(5)$ subgroup of $SO(5,1)$. The spatial embedding coordinates of $AdS_4$, on the other hand, generate the Lie algebra of the non-compact $SO(4,1)$ subgroup of $SO(4,2)$.

The CNC deformation introduces a short distance scale $\ell_p = R/N$, at which space-time coordinates become
fuzzy. Indeed, the main motivation for introducing the deformation is to provide a natural geometric UV cut-off 
for the CFT. For $dS_4$, the number of fuzzy points on the spatial slice is given by the dimension of the $SO(5)$ represention with maximal allowed spin $N$. This suggests that the function space on a given spatial section is truncated at a maximal allowed value of the angular momentum equal to $N$, {\it c.f.} \cite{fuzzyfive}.
In this respect, the CNC deformation indeed seems to act as a UV regulator.
The non-trivial new ingredient of our construction is that this short-distance cut-off preserves Lorentz invariance.

\bigskip

\noindent
{\it Gravitational interactions?}

Our secret motivation for this work is that, by introducing a geometric covariant cut-off,  we are in fact automatically coupling the CFT to gravity, with a finite Planck length proportional to the short distance cut-off $\ell_p$ \cite{us}. There are
three separate pieces of evidence that support this possibility:
\smallskip

{\it Hint 1}: We know from AdS/CFT  that a strongly coupled  3+1-D continuum CFT is dual to gravity on a non-compact asymptotically $AdS_5$ geometry. By introducing a UV cut-off, we are effectively compactifying the $AdS_5$, by removing the non-compact asymptotic region. The cut-off acts like an effective Planck brane, with an associated normalizable 5D graviton zero mode. Assuming that the boundary conditions are such that the zero mode is allowed to fluctuate, it will act as a 4D graviton coupled to the cut-off CFT. 

\smallskip

{\it Hint 2}: Consider a CFT with a covariant UV cutoff. The geometric entanglement entropy associated with a subregion of space is then proportional to a finite constant times the surrounding surface area. Jacobson has convincingly argued that this fact, in combination with Lorentz invariance and the assumption that entanglement entropy satisfies the first law of thermodynamics, is sufficient to conclude that the background geometry must be dynamical and to derive that its dynamics is described by the Einstein equations.  Newton's constant is then normalized such that the entanglement entropy matches with the Bekenstein-Hawking formula. This second hint can be directly linked to the first hint via the Ryu-Takayanagi formula for the holographic entanglement entropy \cite{rt,faulkner}.

\smallskip

{\it Hint 3}:  In this paper, we have computed the leading order correction term to the 2-point function of two scalar CFT operators. The result, given in equation (\ref{twopcnc}), is covariant and proportional to $\ell^2$. It is natural to compare this leading order CNC correction with the leading order graviational correction due to a single graviton exchange
\bea
\kappa^2 \; \int \!\! d^4x \spc d^4y \, \Bigl\langle\spc {\cal O}(X_2) \; \frac{ T_{\mu\nu} (x)\, T^{\mu\nu} (y) }{( x\nspc - \nspc y )^{2}}\; {\cal O}(X_1)\spc  \Bigr\rangle
\eea
While we have not yet been able to explicitly compute this gravitational correction, general physical reasoning (dimensional analysis, Lorentz symmetry, quadratic dependence on conformal dimension $\Delta$) suggests that the answer should look identical to (\ref{twopcnc}) with $\kappa = \ell_p$.

\bigskip

\noindent
{\it Regge trajectory?}

In section 5 we computed the first order CNC correction to the $n$-point function of scalar primary operators, with the result (\ref{ncorr}). The appearance of the polarization tensors in the first order correction indicates that the CNC deformation produces new states with non zero spin. 
How should we interpret these states? Where do they come from? We do not yet have a sufficient understanding of the CNC deformed theory to give a precise answer to these questions, so we will only make some general comments.

The form of the full tower of CNC correction terms suggests that we need to associate to each scalar primary operator ${\cal O}(X)$ of the undeformed CFT, an infinite tower of (non-local) `excited primary operators' with ever increasing spin $j$ and scale dimension $\Delta_j = \Delta + j$, analogous to Vasiliev theory \cite{vasiliev} or a Regge trajectory of excited string states. The whole tower can be packaged in a single operator ${\cal O}(Z)$ defined on the extended space-time, via\footnote{Here $P^\AA$ is the polarization vector related to the $S_\AB$ tensor via  $S^\AB = P^{[\AA} X^{\BB]}$.} 
\bea
\label{reggesum}
{\cal O}(X)  & \to &   {\cal O}(Z) \, =\, {\cal O}(X)\, +\, P^\AA \spc {\cal O}_{\AA}^{(1)} (X) \,+\, P^\AA P^\BB {\cal O}_{\AB}^{(2)} (X)\; + \, \ldots,
\eea
Like excited string states,  the excited operators are decoupled in the commutative limit, but can contribute as soon as the CNC deformation is turned on.  However, unlike excited string states, 
they not only contribute in factorization channels, but also appear to modify the
asymptotic states themselves. From equation (\ref{ncorr}) and (\ref{reggesum}), we find that
the contribution of the first exited state of, say, the $k$-th operator ${\cal O}_k(X_k)$ takes the form 
\bea
\Bigl\langle\spc {\cal O}_n(X_n) \ldots {\cal O}^{(1)}_A(X_k) \ldots  \; {\cal O}(X_1)\spc  \Bigr\rangle
\, = \, \qquad \qquad \qquad\qquad\qquad\qquad\qquad \qquad \qquad \qquad\nonumber\\[-2mm]\\[-2mm]\nonumber
 \frac{\hbar \ell^2}2\spc 
 \sum_{i} 
 {{\rm sgn}(i\!-\! k)} 
 \Bigl(\bigl(\nspc X_i\!\nspc\cdot\! \partial_i
\bigr) \, \bigl(\partial_{k}\bigr)_{\! \AA}\!
- \bigl(\nspc X_k\!\nspc\cdot\! \partial_k \bigr) \bigl(\nspc \Lambda_{ki} \!\nspc \cdot \nspc \partial_i\bigr)_{\! \AA}\Bigr)\Bigl\langle\spc {\cal O}_n(X_n)\, ...\, {\cal O}(X_k) \, ... \, {\cal O}(X_1)\spc  \Bigr\rangle
\eea
with $\Lambda_{ki}$ the $SO(4,1)$ isometry that rotates $X_i$ to $X_k$.
We explicitly see that the CNC deformation breaks crossing symmetry. In other words, the excited states can not be viewed as descendant CFT operators, but as new operators without any local precursors in the CFT.
Note that the first excited operator ${\cal O}^{(1)}_A(X_k)$ indeed has scale dimension $\Delta_k + 1$.
\begin{figure}[t]
\begin{center}
\vspace{-.8cm}

\includegraphics[scale=.58]{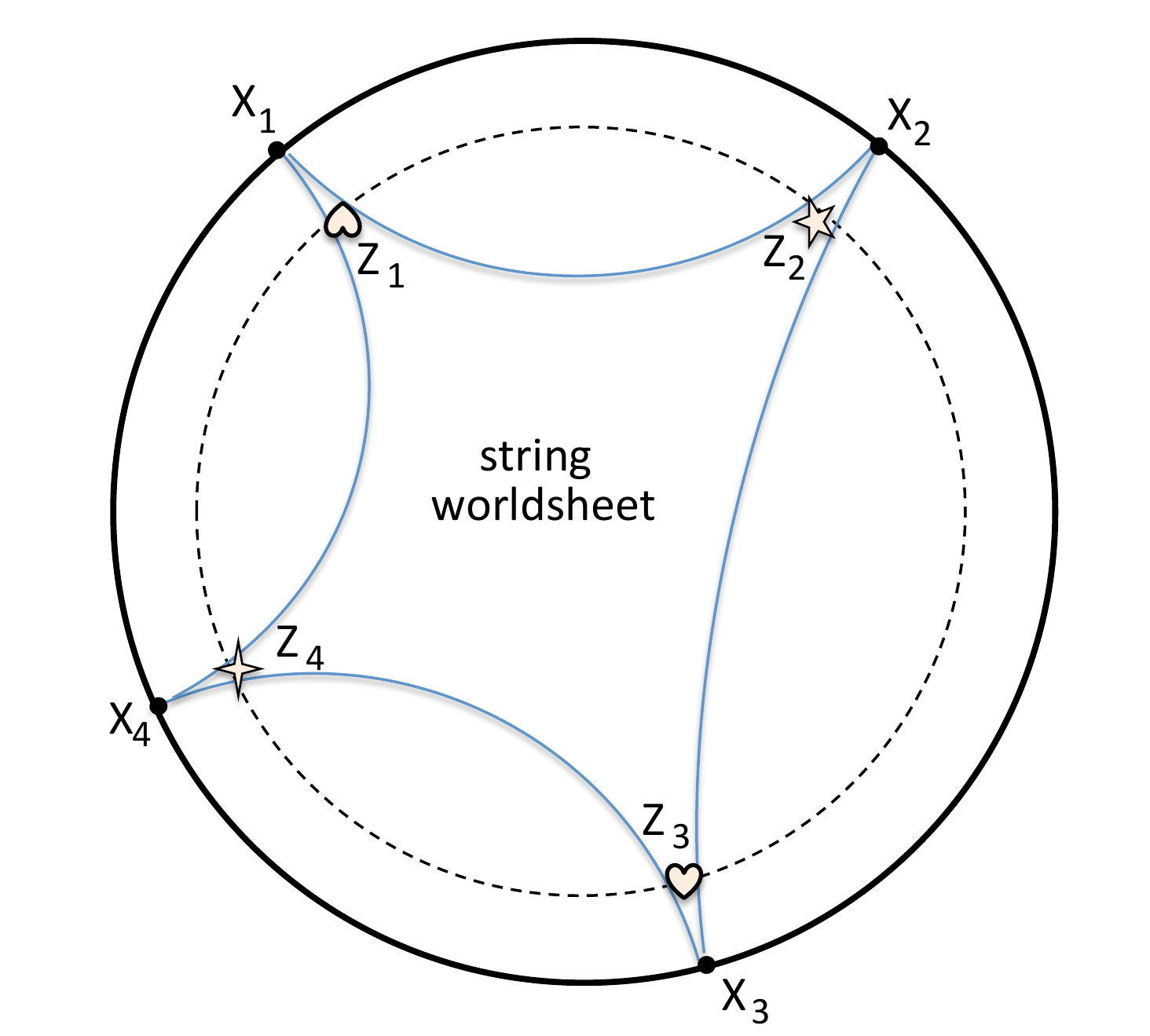}
\caption{\small An intuitive explanation for the appearance of higher spin states. A continuum CFT correlator corresponds to an on-shell string amplitude between point like string states that reach the asymptotic AdS boundary. In the deformed theory, the asymptotic region is removed. The strings maintain a minimal size and remain polarized. This leads to higher spin excitations.
 }
\end{center}
\vspace{-0.5cm}
\end{figure} 

  A rough intuitive explanation for the appearance of the higer spin degrees of freedom is shown in figure 1.
It indicates how the CNC deformation modifies the continuum CFT correlator, from an AdS dual perspective.  The deformed theory has a minimal length, which effectively removes the asymptotic region of AdS. The string scattering states thus no longer have the possibility to go fully on-shell: they have a minimal size proportional to $\ell_p$ and thus retain part of their shape information in the form of additional spin degrees of freedom. There is indeed some similarity between our CNC deformation and the use of
star products in the construction of Vasiliev's higher spin theories \cite{vasiliev}. The partial breaking of crossing symmetry suggests that the bulk strings are polarized into open strings.

\def\cX{{X}}
\def\bcX{\,\overline{\!{X}}}
\def\cN{N} 

\def\bA{\spc \overline{\! \AA}}

\bigskip
\bigskip
\bigskip

\noindent
{\large \bf Acknowledgements}

We thank C. C\'{o}rdova, S. Giombi, T. Hartman, D.M Hofman, T. Jacobson, J. Maldacena, D.
Skinner, C. Vafa, E. Verlinde and E. Witten for helpful discussions. The work
of JJH is supported by NSF grant PHY-1067976. The work of HV is supported by
NSF grant PHY-1314198.


\bigskip
\bigskip
\bigskip

\noindent
{\large \bf Appendix A: CBH Star Product}

In this Appendix we define the symbol $D(Z,a,b)$ that appears in the $SO(5,1)$ invariant star product (\ref{dstar}).
Define two elements $A$ and $B$ of the $so(5,1)$ Lie algebra
\bea
\label{abdef}
A \spc =\spc Z_\IJ \, a^\IJ, \quad & & \quad B\spc =\spc Z_\IJ\, b^\IJ
\eea
where $Z_\IJ$ denote the non-commutative coordinates, satsifying the commutation relations (\ref{zcom}). 
Now consider the CBH formula
\bea
\exp{A}\! \!\!\!\!\!\!\!\!\!\!\! &  & \cdot \exp{B} \, =\, \exp\bigl({A+B + D(A,B) }\bigr) \\[5mm]
D(A,B)  \is \frac 1 2 \bigl[A,B\bigr] + \frac 1 {12} \Bigl(\bigl[ A, [A,B]\bigr] + \bigl[[A,B],B\bigr]\Bigr) + \ldots
\label{dexp}
\eea
Upon inserting the component expansion (\ref{abdef}) of $A$ and $B$, and using the commutator algebra (\ref{zcom}),
we can expand $D(A,B)$ as a linear function of $Z_\IJ$
\bea
D(A,B) = Z^\IJ D_\IJ(a,b) 
\eea
This uniquely defines the symbol $D_\IJ(a,b)$ used in the star product (\ref{dstar}).

\addtolength{\baselineskip}{-.3mm}

\end{document}